\documentclass[manuscript]{aastex}
\usepackage{graphicx}  
\usepackage{gensymb}
\usepackage{float}

\begin{document}

\title{Observational Characteristics of CMEs without Low Coronal Signatures}

\author{E. D'Huys\altaffilmark{1,2}}
\email{elke.dhuys@observatory.be}
\author{D. B. Seaton\altaffilmark{1}}
\author{S. Poedts\altaffilmark{2}}
\and
\author{D. Berghmans\altaffilmark{1}}
\altaffiltext{1}{Royal Observatory of Belgium,  Ringlaan -3- Av. Circulaire, 1180 Brussels, Belgium}
\altaffiltext{2}{Katholieke Universiteit Leuven, Centre for mathematical Plasma-Astrophysics, Ce\-les\-tij\-nen\-laan 200b - bus 2400, 3001 Leuven, Belgium}

\begin{abstract}
Solar eruptions are usually associated with a variety of phenomena occurring in the low corona before, during, and after onset of eruption. Though easily visible in coronagraph observations, so-called \textit{stealth} coronal mass ejections (CMEs) do not obviously exhibit any of these low-coronal signatures. The presence or absence of distinct low coronal signatures can be linked to different theoretical models to establish the mechanisms by which the eruption is initiated and driven. In this study, 40 CMEs without low coronal signatures, occurring in 2012, are identified. Their observational and kinematic properties are analyzed and compared to those of regular CMEs. 

Solar eruptions without clear on-disk or low coronal signatures can lead to unexpected space weather impacts, since many early warning signs for significant space weather activity are not present in these events. A better understanding of their initiation mechanism(s) will considerably improve the ability to predict such space weather events.
\end{abstract}



\section{Introduction}

Coronal mass ejections (CMEs) are observed as bright transient features, suddenly appearing in white-light coronagraph observations. Even if their exact relationship to eruptive events in the low corona remains a matter of debate, these CMEs are very frequently accompanied by eruptive or dynamical phenomena low in the solar atmosphere: solar flares, flows, magnetic reconfiguration, EUV waves, jets, coronal dimmings or brightenings, filament eruptions, or the formation of post-flare loop arcades. However, coronal mass ejections that cannot be associated with any of these low coronal signatures (LCS) of eruption have been observed as well. This lack of association makes it difficult to determine their solar source region, which, in turn, makes them difficult for space weather forecasters  to assess and has earned them the title \textit{stealth} CMEs.  A seminal case-study of a stealth CME was published by \citet{Robbrecht2009_stealth}. This publication described a streamer-blowout CME without obvious EUV and H$\alpha$ signatures that apparently originated high in the solar corona, thus explaining the lack of on-disk signatures. 

Stealth CMEs appear to be less uncommon than the low number of published case studies suggests. Studying the source locations of all 1078 CMEs listed in the CDAW CME catalog\footnote{http://cdaw.gsfc.nasa.gov/CME\_list/} during 1997-1998, \citet{Wang2011} found a considerable number of events ($\sim16 \%$) that were assumed to be front-sided, but lacked eruptive signatures in the EIT 19.5~nm images. 

\citet{Ma2010} carried out a statistical study of CMEs without distinct low coronal signatures. Their dataset spanned the period from January 1 to August 31, 2009, which was a time of an exceptionally low solar minimum. They report that almost one out of three CMEs in their catalog turned out to be stealth and that nearly half of the CMEs without LCS was a blowout type CME. A kinematic study of the 11 identified stealth CMEs revealed that these were slow CMEs ($v < 300$ $\textrm{km s}^{-1}$) that were accelerated gradually and had an angular width smaller than $40\degree$.  

\citet{Howard2013} point out that while the paper by \citet{Robbrecht2009_stealth} gave rise to the term \textit{stealth CME} in several subsequent publications, the concept of so-called \textit{problem storms} is found much earlier in the literature, referring to geomagnetic storms without an obvious solar counterpart. As a result, the terms \textit{problem storms} and \textit{stealth CMEs} are sometimes used interchangeably. We advise careful wording, however, since the former applies to geomagnetic effects observed near Earth, while the latter refers to the solar origin of these space weather effects. Moreover, many stealth CMEs are not earth-directed and thus do not cause a problem storm.

The central question driving the research presented here is whether CMEs without low coronal signatures are fundamentally different from other CMEs. Do both classes of CMEs have different initiation and driving mechanisms or are CMEs without LCS simply at the low end of an energy spectrum, making their associated surface signatures hard to observe? Indeed, as \citet{Howard2013} point out, one needs to keep in mind that detections of eruption signatures are limited by the sensitivity and bandwidth of the instrumentation used. 

There is no agreement within the solar physics community on the definition of a stealth CME. \citet{Ma2010} define a CME without low coronal signatures, where LCS means a "filament eruption, flare, post-eruptive arcade, coronal wave, coronal dimming, or jet". Alternatively, \citet{Wang2011} specify  "a kind of CME that does not leave any eruptive signatures in EUV-passbands and sometimes may not even be visible in coronagraphs facing on them". Notice, however, that the prime example of a so-called stealth CME, the one studied by \citet{Robbrecht2009_stealth}, does not fit this last definition, as a careful examination of EUVI-A 17.1~nm images for this event revealed a bright structure at $0.15~R_\odot$, travelling outward to form the CME core. Thus EUV images did show an eruptive signature for this event, albeit at a large height. This is also the case for most events studied by \citet{Ma2010}. They report that 8 out of 11 identified CMEs without LCS may be initiated by disturbances of flux ropes suspended high in the corona.

For the purpose of this study, we have defined a CME without low coronal signatures as a front-sided CME that was detected in coronagraph images and for which no coronal signature was observed on the solar disk or in the more extended field-of-view of the EUV-imagers PROBA2/SWAP \citep{Seaton2013}, SDO/AIA  \citep{Lemen2012} and STEREO/EUVI  \citep{Howard2008}. This definition introduces a clearer distinction between stealth CMEs and other events than the definitions listed above. Indeed, what makes stealth CMEs stand out from other events, is exactly the fact that it is very difficult to determine their source regions. In case an eruptive signature is detected at larger height, this would be a clear indication of the origin of the CME and therefore we do not label it as a stealth event. 

\begin{figure}[H]        
\centerline{\includegraphics[height=0.95\textwidth]{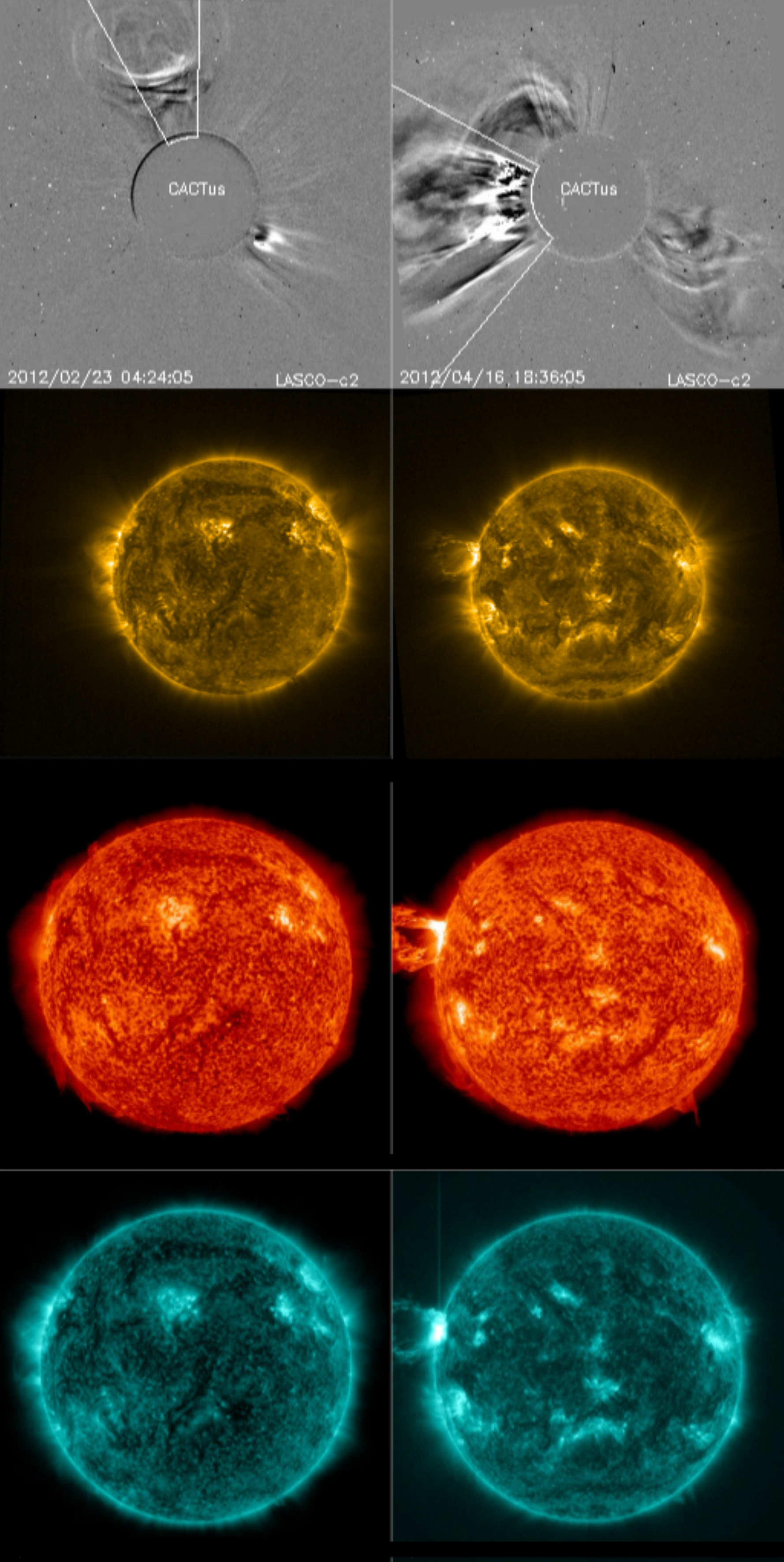}}
\caption{CACTus LASCO detections of a stealth CME (left on the top row) and of a CME associated with coronal signatures (top row, right). Subsequent rows illustrate the coronal signatures associated with these events in different wavelengths (PROBA2/SWAP 174, SDO/AIA 304 and SDO/AIA 131, respectively). The CME on the right was associated with an erupting filament and an M1.7 flare on the east solar limb.}
\label{stealthvsnormal} 
\end{figure}

To classify the events in our dataset into the categories of stealth and non-stealth CMEs according to our definition above, we have searched for possibly related flares and brightenings, filaments, EUV waves, jets, coronal dimmings, flows, post-flare loops, and reconfiguration of the magnetic field lines in the higher corona. Figure~\ref{stealthvsnormal} illustrates the vast difference in low coronal signatures between a stealth CME and a CME associated with a filament eruption and an M1.7 flare.

\section{Searching for CMEs without LCS}

Identifying stealth CMEs is not a straightforward task. In fact, we are looking for a negative association---CMEs that cannot be associated with any low coronal signature---which is much harder than proving a positive one. Our purpose was not to confirm every single candidate stealth CME in our dataset, but rather to find a  number of interesting events to be used for the observational characterization of stealth CMEs and for numerical modeling of specific events. Therefore we used a rather exclusive approach in determining whether a CME could be associated with low coronal signatures or not. 

To eliminate the obvious non-stealth CMEs from the vast dataset we studied, we developed a procedure that combines the output of different automated tools, each one detecting a different low coronal signature of solar eruptions. The input for this algorithm is the list of CME detections produced by CACTus \citep[www.sidc.be/cactus]{Robbrecht2004, Robbrecht2009_cactus}, a software tool that autonomously detects CMEs in the SOHO/LASCO coronagraph images \citep{Brueckner1995}. For each entry, the catalog lists the CME timing information, its principal angle, angular width and median velocity. In this study we focus on the year 2012, during which CACTus detected 1596 CMEs in the LASCO images. (Table~\ref{matchestable})

\begin{deluxetable}{lcc}
\tablecolumns{3}
\tablewidth{0pc}
\tablecaption{Number of matches when comparing CACTus LASCO CME detections to GOES flare lists, COR2 CME observations and SoFAST detections. Our algorithm generated a list of 481 stealth CME candidates, which was reduced to 40 confirmed CMEs without LCS after visual inspection of all these events.}
\tablehead{
 \colhead{Catalog} & & \colhead{Number of Matches}
 }
\startdata
CACTus CME list & & 1596 \\
\hline
GOES event list & & 680 \\
CACTus COR2-A CME list & & 396 \\
CACTus COR2-B CME list & & 413 \\
SoFAST catalog & & 332 \\
\hline
Stealth CME Candidates & & 481 \\
\hline
Confirmed Stealth CMEs & & 40 \\
\enddata
\label{matchestable}
\end{deluxetable}

The CACTus CME catalog for 2012 was coupled with the Geostationary Operational Environmental Satellite X-ray event lists \citep[GOES/XRS,][]{Hanser1996} to filter out CMEs with an associated X-ray flare. We associate an X-ray flare to a CME in cases where the flare occurred at most 4000~s before the initial detection of the CME by CACTus, or when it was observed less than 3600~s after. These are empirically derived time limits that were found to result in the best matches between associated events. The permitted time interval between a flare before a CME and the CME itself was further adjusted according to the CME speed as measured by CACTus. This adjustment was based on the kinematics of a particle trajectory under constant acceleration, with an upper limit fixed at 4000~s. As a result,  680 out of 1596 LASCO CMEs were matched to a GOES X-ray flare. This is illustrated in Table~\ref{matchestable}.

Next, the algorithm compared the CACTus LASCO CME catalog to the CACTus CME detections in  SECCHI/COR2 coronagraph images \citep{Howard2008} to exclude back-sided CMEs. A CACTus LASCO CME was identified as back-sided in case an  associated CACTus COR2 event was found that occurred within one hour either side of the CACTus detection time in LASCO and for which the COR2 principal angle indicated the CME was propagating away from Earth. On June 30, 2012, the midpoint in the time period that was investigated, the separation angle between STEREO A and Earth was around $119\degree$, while the separation between STEREO-B and Earth reached $116\degree$, implying that by combining these three viewpoints the complete solar surface could be observed. For the purpose of this study, back-sided CMEs were coarsely defined as having a principal angle (counterclockwise) in the range of $180\degree$ to $360\degree$ in the case of COR2-A, and a principal angle between $0\degree$ and $180\degree$ for COR2-B observations. Accordingly, 396 LASCO CMEs were determined to occur on the far side of the sun based on COR2-A data, while 413 events were back-sided as seen from COR2-B. (Table~\ref{matchestable})

Finally, the CACTus LASCO list was compared to the output of the \textit{Solar Flare Automated Search Tool} \citep[SoFAST,][www.sidc.be/sofast]{Bonte2013}, based on observations from PROBA2/SWAP. SoFAST allows for the elimination of events with any associated EUV variability. Table~\ref{matchestable} shows that 332 LASCO CMEs were found to be connected to variability in the SWAP images. 

Each of these steps was performed independently and in the case that a specific CME from the CACTus LASCO catalog was associated to any of the detections in the other datasets, it was removed from the list of candidate stealth CMEs. Applying this procedure to the data for the year 2012 resulted in a list of 481 CMEs that could not be linked automatically with flares, EUV brightenings or activity on the far side of the sun. (Table~\ref{matchestable}) Visual inspection of solar images in various wavelengths using observations from PROBA2/SWAP, SDO/AIA and STEREO/SECCHI for all these events enabled us to eliminate CMEs associated with filament eruptions, EUV waves or dimmings, or eruptive signatures at larger heights. Some events could also be linked to flares or back-sided CMEs occurring outside the time intervals that we implemented to exclude events from the CACTus LASCO CME list. This final effort resulted in a list of 40 confirmed CMEs without low coronal signatures, displayed in Table~\ref{stealthtable}.

At this point it is important to emphasize once more that this procedure was not designed to extract every single CME without low coronal signatures that occurred in 2012 directly from the input catalog. The purpose was instead to find a sufficiently large number of interesting stealth events to investigate in more detail. The algorithm was developed to eliminate as many CMEs with clear observational signatures as possible following an automated procedure, thus limiting the number of  events remaining for visual inspection. 

Undoubtedly, during this procedure a limited number of incorrect associations was made between a CME and the detection of a flare, EUV variability or a CME on the far side of the sun, mainly because these associations were based on timing only: information on CME principal angles and flare locations was ignored. In order to assess the algorithm's performance, we used the location information provided in the GOES event list and in the SoFAST catalog to approximate the principal angle of propagation for CMEs related to these events and the resulting principal angles were compared to those of the matching CACTus CMEs. The  principal angles calculated by CACTus for LASCO and COR2 CMEs that were paired up by the algorithm were correlated as well. This procedure is subject to certain limitations. For example, when calculating the  principal angle for SoFAST and GOES events, we are assuming that the associated CME is propagating radially outward from its source region, while in fact it may undergo a considerable deflection \citep[e.g.][]{Zuccarello2012}. However, the  principal angles are found to agree reasonably well, taking into account these constraints. 

Despite its limitations, this procedure allowed us to sift through the large number of detections in the 2012 CACTus LASCO CME catalog in an objective, automated and reproducible manner. As a result, we were able to confirm the occurrence of 40 CMEs that indisputably had no low coronal signatures. To our knowledge, this is the largest sample of stealth CMEs studied so far. 


\section{Observational properties of CMEs without Low Coronal Signatures}

We used the 40 identified CMEs without low coronal signatures and their corresponding CACTus LASCO detections shown in Table~\ref{stealthtable} to characterize the general properties of stealth CMEs. CME appearance, position angle, velocity, and angular width were studied and compared to those of CMEs with LCS. We also studied the scale invariance of stealth CMEs. 

When interpreting these results, it is important to remember that our sample of CMEs without low coronal signatures is limited to 40 events, a low number compared to the nearly 1600 events in the complete CACTus LASCO catalog for 2012.

\subsection{Appearance in coronagraph images}

\begin{figure}[H]    
\centerline{\includegraphics[width=1.0\textwidth]{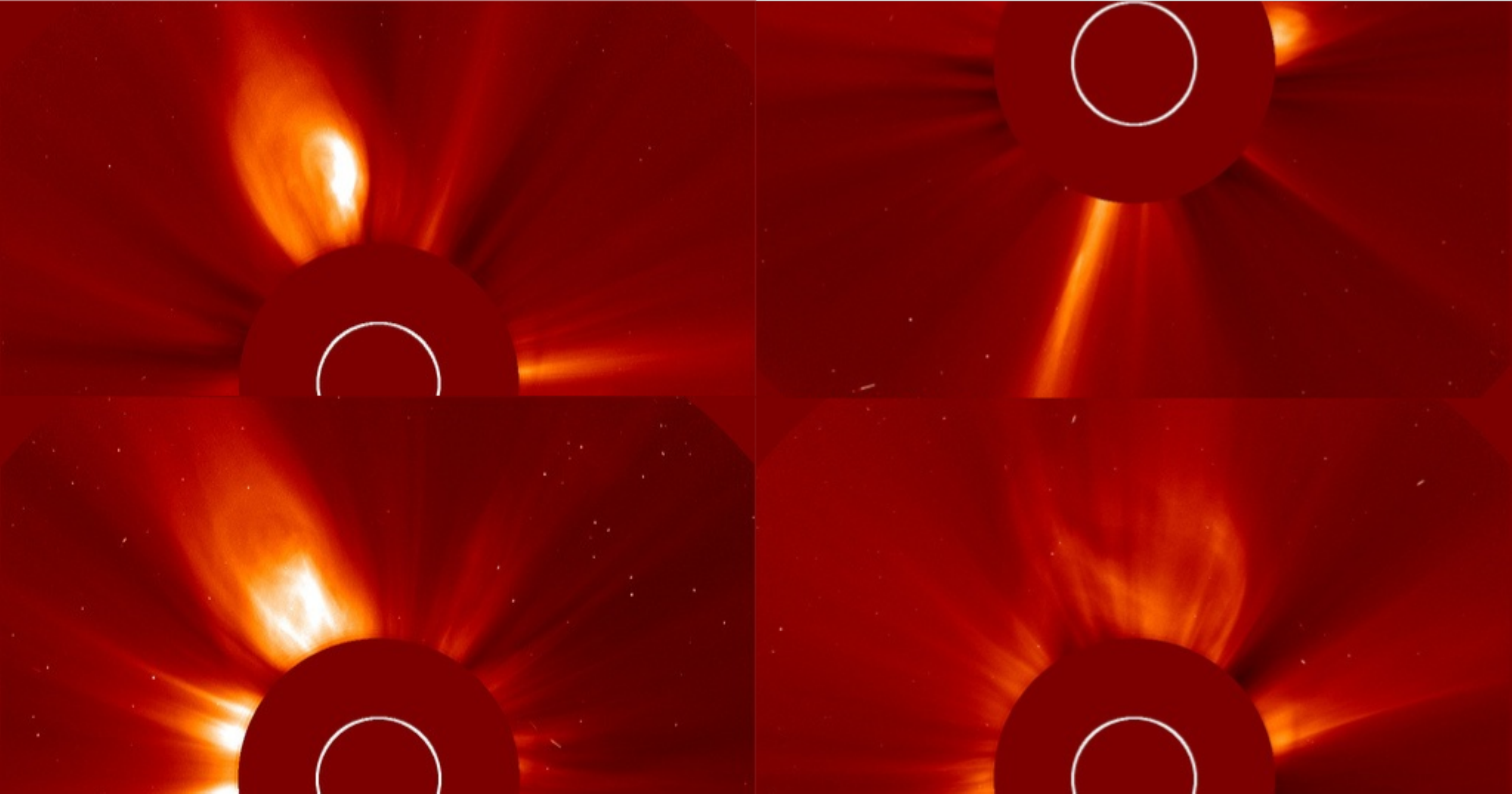}}
\caption{LASCO coronagraph observations of CMEs without low coronal signatures illustrating the variety in appearance for these events. Shown in clockwise direction, starting in the upper left corner, are a three-part CME, a narrow CME, a wide CME (angular width of 80\degree, measured by CACTus) and a streamer blowout CME.}
\label{appearance} 
\end{figure}

Observationally there is a large diversity in the appearance of CMEs without low coronal signatures in coronagraph images. Some events are very narrow and similar to outflows while others are very wide CMEs, in some cases also showing the typical three-part structure. A few examples are shown in Figure~\ref{appearance}.  Seven out of 40 CMEs without low coronal signatures were of the streamer blowout type. 

In 73\% of events, the stealth CME was preceded by another nearby CME, which could conceivably have destabilized the overlying magnetic field and thus contributed to the initiation of the stealth event. Many of the stealth CMEs occurred in the vicinity of the polar coronal holes, a region of largely open magnetic field, another factor that may have contributed to facilitating a stealth eruption. Indeed, if an eruption occurs in a region where overlying field provides very little downward directed force on the erupting structure, that is an open field region, it may be that the eruption can unfold without a major restructuring of the magnetic field and therefore no strong low coronal signatures of eruption are observed. For the CMEs without LCS that had a presumed source region closer to the equator, the PFSS reconstructions also showed open field lines nearby for four out of five events.

\subsection{Position angle}
\begin{figure}[H]   
\centerline{\includegraphics[width=0.8\textwidth]{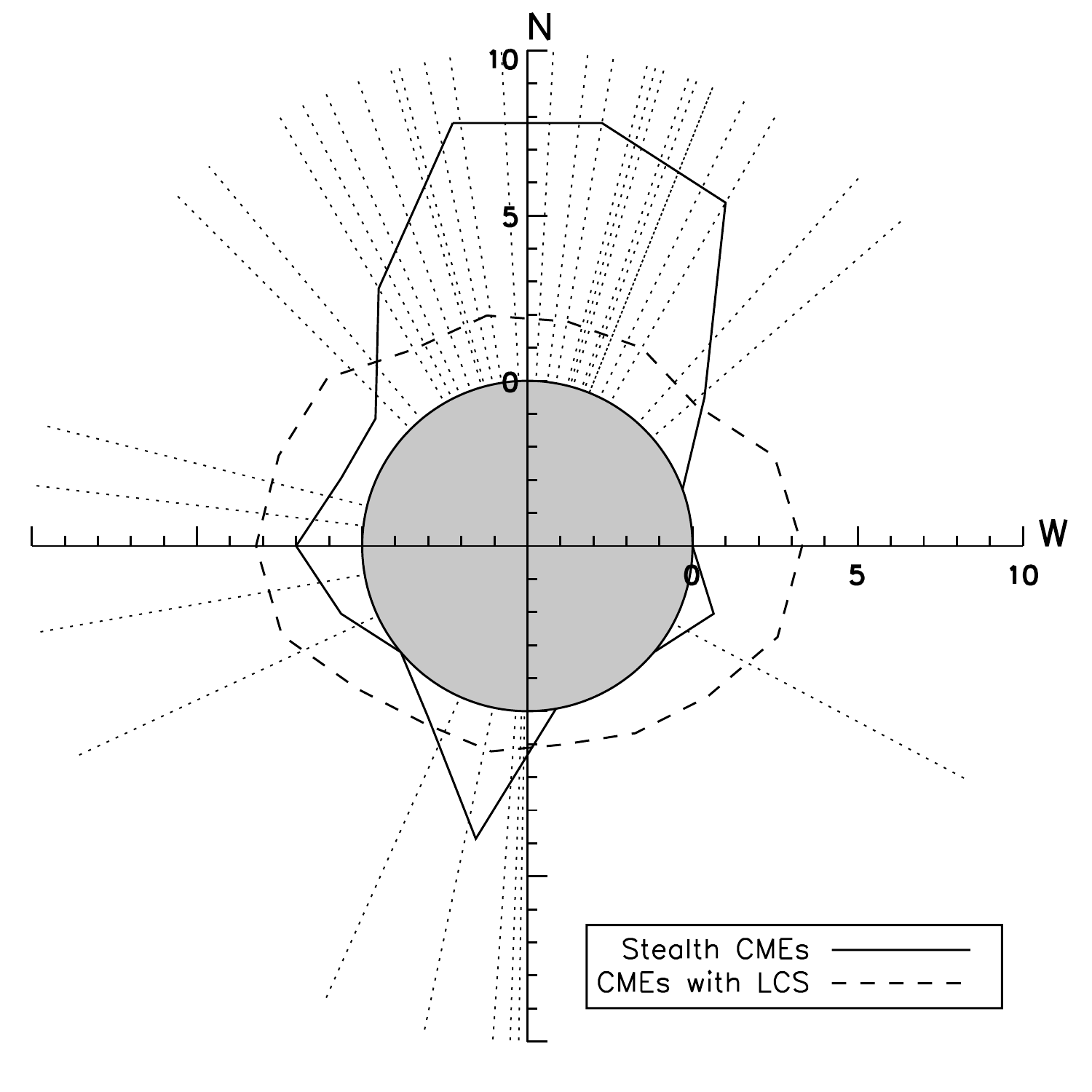}}
\caption{Distribution of coronal mass ejections with (dashed line) and without (full line) low coronal signatures relative to their principal angle of propagation, binned according to their principal angle with a bin size of 20\degree . The number of occurrences of CMEs with coronal signatures was scaled down proportionally by a factor $(1596-40)/40$. The dotted lines indicate the principal angle of propagation for each stealth CME as measured by CACTus.}
\label{polar_plot} 
\end{figure}

It is striking that many of the events in our list of CMEs without LCS have a principal angle directed towards the north. This is illustrated in Figure~\ref{polar_plot}, where the distributions of CMEs with and without LCS are plotted as a function of the principal angle measured by CACTus. Note that, for ease of comparison and to allow for the plotting of both curves on the same axes, the number of occurrences of CMEs with coronal signatures was scaled down proportionally by a factor of $(1596-40)/40$. Evidently, CMEs that exhibit low coronal signatures of an eruption are much more evenly spread across the solar disk than stealth CMEs are. The fact that many CMEs without LCS seem to originate at high northern latitudes and near the polar coronal hole, suggests that their source region is not a magnetically complex region, which is compatible with the lack of coronal signatures and the low speeds (see below) of these coronal mass ejections. 

It remains important to emphasize that these findings are based on a small number of stealth events. However, when a random set of 40 events is taken from the CACTus CME list for 2012, the principal angle distribution is in a large majority of the cases randomly spread around the solar disk. Only eight out of 1000 random samples of 40 CMEs (i.e. less than $1\%$) had at least 20 events directed towards the north, where an event towards the north was defined as having a principal angle that fell between $300\degree$ and $60\degree$, with north corresponding to an angle of $0\degree$. Thirty out of the 40 CMEs without low coronal signatures studied here, fit that definition. This clearly illustrates that the predominantly northward propagation of our sample of stealth events is not just a stochastical coincidence, but an inherent property of the CMEs without low coronal signatures studied here.

We investigated the possibility that this apparent preference for stealth CMEs to originate near the north pole was an observational effect caused by the tilt of the solar rotation axis. Indeed, when the solar north pole is directed towards earth, possible low coronal signatures in this region should be easier to observe and thus it should be more straightforward to determine whether a CME is stealth or not. Additionally, more stealth CMEs originating near the northern pole would be labeled as front-sided events. However, no clear relationship between the solar $\mathbf{ B_0}$ angle, characterising the tilt of the solar rotation axis with respect to the ecliptic north, and the number of stealth CMEs towards the north (or south) could be identified.

Another potentially contributing factor is the dominance of the northern hemisphere in solar activity during the year 2012. This can clearly be seen, for example, from the hemispheric sunspot numbers during this period\footnote{http://sidc.oma.be/silso/monthlyhemisphericplot}. Far more sunspots and active regions were observed above the solar equator than below. This explains the slight imbalance in the spread across the solar disk of the CMEs with LCS: more events were detected with a principal angle pointing towards the north. The same effect is expected for stealth CMEs. However, that observation alone is probably not sufficient to explain the large discrepancy in northward and southward directed stealth CMEs that is apparent in Figure~\ref{polar_plot}.

\subsection{CME speed and velocity profiles}

\subsubsection{Velocity Distribution}
\begin{figure}[H] 
\centerline{\includegraphics[width=1.0\textwidth]{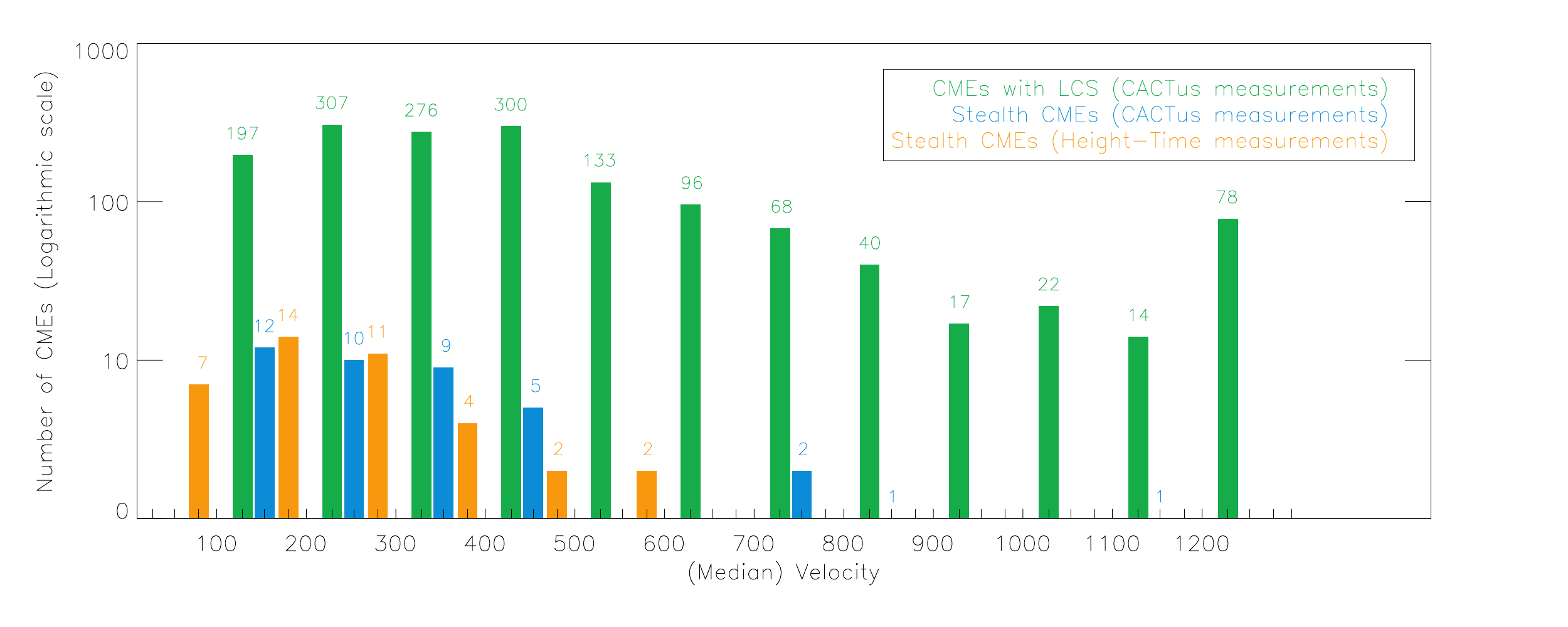}}
\caption{Distribution of the median velocity for coronal mass ejections with and without low coronal signatures as measured by the CACTus algorithm and plotted on a logarithmic scale (bin size = 100 km/s). The third distribution represents the projected CME velocities based on height-time measurements of the bright leading edge of the CME in the LASCO images.}
\label{medianvelocity} 
\end{figure}

The CACTus CME detection algorithm reports the median velocity for each observed CME. The software determines the speed of the CME in each direction within the angular span of the CME. The median of the resulting velocity profile is given as the speed of the CME. The distributions of these median velocities for CMEs with and without low coronal signatures are shown in Figure~\ref{medianvelocity} on a logarithmic scale. The median speeds calculated by the CACTus software tend to differ from the velocities of CMEs reported in other (manual) catalogues, mostly because the latter are usually based on measurements of the bright leading edge. \citep{Robbrecht2004} For comparison with these catalogues, the stealth CME velocity, along the principal angle and projected in the plane of the sky, was derived from height-time profiles of the bright leading edge observed in the LASCO images. The resulting distribution is shown in Figure~\ref{medianvelocity} as well. 

Figure~\ref{medianvelocity} clearly illustrates that the peak of the velocity distribution occurs at lower values for stealth CMEs compared to CMEs with LCS.  It follows that CMEs without low coronal signatures are generally slow events: typically with a median velocity between 100~$\textrm{km s}^{-1}$ and 500~$\textrm{km s}^{-1}$, although a few faster eruptions were also identified. This low velocity is not surprising as the lack of on-disk signatures suggest there may only be limited free energy available, not enough to drive a very fast eruption. 

Our results are compatible with the distributions obtained by \citet{Ma2010}. These authors found 11 stealth CMEs with velocities ranging from 100~$\textrm{km s}^{-1}$ to 300~$\textrm{km s}^{-1}$. The fact that the present study also identified stealth CMEs with higher velocities could simply be linked to the larger number of stealth CMEs found here. An additional influence may come from the different phases of the solar cycle in which the CMEs in both surveys were detected. \citet{Ma2010} investigated stealth CMEs in the first half of 2009, a period of deep solar minimum, while during 2012  solar activity had increased, rising towards a new solar maximum. Indeed, \citet{Yashiro2004} studied the properties of CMEs observed in the LASCO coronagraph between 1996 and 2002, and found that their average speed increased from 300~$\textrm{km s}^{-1}$ at solar minimum to 500~$\textrm{km s}^{-1}$ at the time of solar maximum.

\begin{figure}[H] 
\centerline{\includegraphics[width=1.0\textwidth]{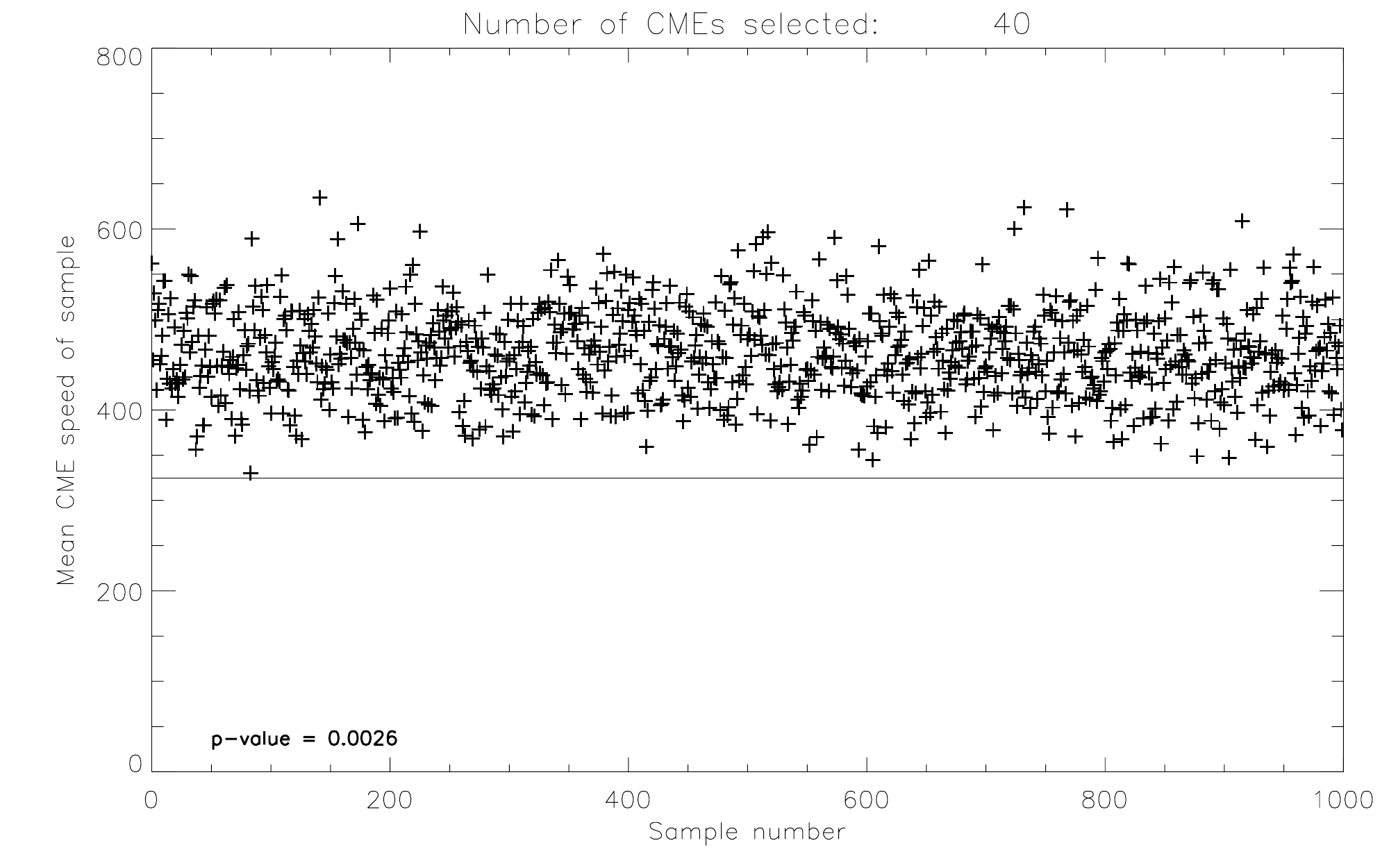}}
\caption{Mean CME velocity of 1000 samples of 40 randomly selected CMEs. The horizontal line indicates the mean velocity of the set of CMEs without low coronal signatures.}
\label{samplevelocity} 
\end{figure}

To assess the influence of our small sample size on the stealth CME velocity distribution, we computed the mean velocity for 1000 samples of 40 CMEs, randomly selected  out of the CACTus detection list for 2012. These values are shown in Figure~\ref{samplevelocity}. This figure clearly illustrates that the mean CME velocity for CMEs without low coronal signatures is much lower. The mean velocity values for the random samples are consistent with a normal distribution with mean $\mu = 461.12$ and standard deviation $\sigma = 48.80$. The mean of the stealth CME velocities is $324.48$ $\textrm{km s}^{-1}$. The probability to obtain this value from the gaussian distribution formed by the means of the random samples is as low as $0.3\%$ ($p = 0.0026$), implying that the group of stealth CMEs indeed stands out from a random set of 40 events.

\subsubsection{Height-time diagrams and velocity profiles}

We compared the height-time evolution of stealth CMEs to published results for different eruption mechanisms \citep[see][for example]{Schrijver2008}. These authors compared filament rise profiles to results from numerical simulations in order to constrain the mechanisms by which the flux rope was destabilized. For example, in case of the two-dimensional catastrophe model by \citet{Priest2002}, the height-time profile takes the form of a power-law with an exponent around 2.5. An exponential rise is compatible with the kink instability \citep{Torok2004, Torok2005} and also with  the torus instability \citep{Kliem2006}, which in fact starts as a $\sinh(t)$ function, and thus is very similar to the exponential function. A parabolic profile is a good description for the CME rising phase in the breakout model \citep{Lynch2004}.

\begin{figure}[H]    
\centerline{\includegraphics[width=0.8\textwidth]{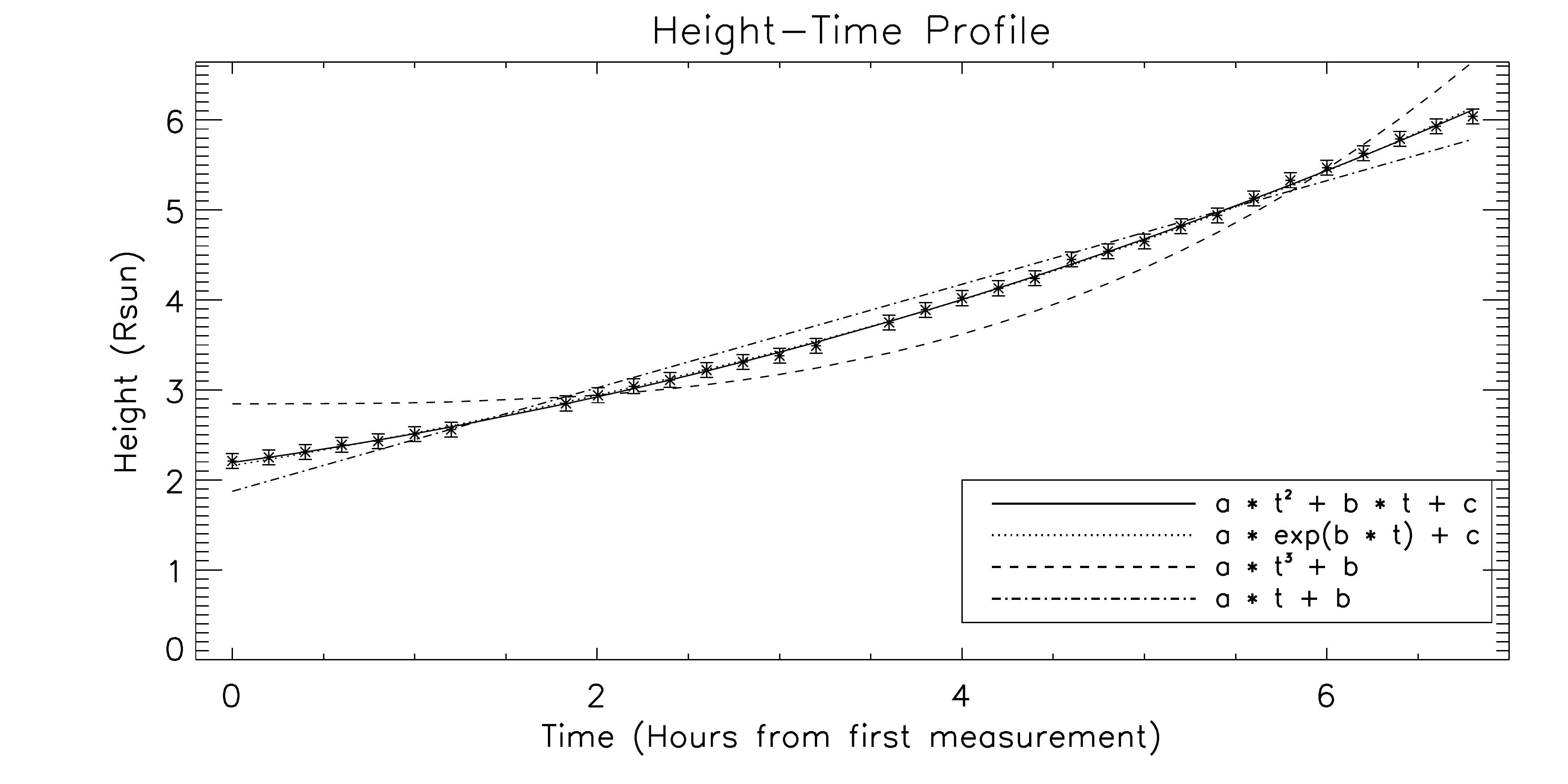}}
\caption{Height-time diagram for a stealth CME observed on February 22, 2012 at 23:48 UT by SOHO/LASCO. (This event is also shown in the left panels of Figure~\ref{stealthvsnormal}.) The height-time measurements were fitted with a parabolic, exponential, cubic and linear function (solid, dotted, dashed and dot-dashed line respectively). The best fits are found for the exponential and parabolic functions. The parameter values for these fits are given in Table~\ref{parameters}.}
\label{heighttime} 
\end{figure}

The best fits for our measurements are exponential and parabolic profiles, corresponding to ideal MHD instabilities and breakout, respectively. An example is shown in Figure~\ref{heighttime} and the parameter values for these fits are given in Table~\ref{parameters}. The lack of LCS suggests that these eruptions are indeed not driven by impulsive reconnection near the solar surface, which is consistent with the evidence from our height-time profiles.

\begin{deluxetable}{lccc}
\tablecolumns{4}
\tablewidth{0pc}
\tablecaption{Parameter values for the best fits to the height-time profile of the stealth CME observed on February 22, 2012 at 23:48 UT by SOHO/LASCO. In the last column the reduced $\chi^2$ value is given, a goodness-of-fit parameter that takes the number of datapoints into account and normalizes for the model complexity. The closer this value is to 1, the better the model fits the observations.}
\tablehead{
 \colhead{Profile} & \colhead{Parameter} & \colhead{Value} & \colhead{Reduced $\chi^2$}
 }
\startdata
$a t^2 + b t + c$ & a & 0.044 & 0.107 \\
 & b &  0.276 & \\
  & c &  2.194 & \\
\hline
$a  \exp(b t) +  c$ & a &  2.224 & 0.173 \\
 & b &  0.151 & \\
  & c &  -0.066 & \\
\hline
$a  t^3 + b $ & a & 0.012 & 18.000 \\
 & b &  2.846 & \\
\hline
$a  t + b $& a & 0.576 & 4.370 \\
 & b &  1.873 & \\
\enddata
\label{parameters}
\end{deluxetable}

\begin{figure}[H] 
\centerline{\includegraphics[width=1.0\textwidth]{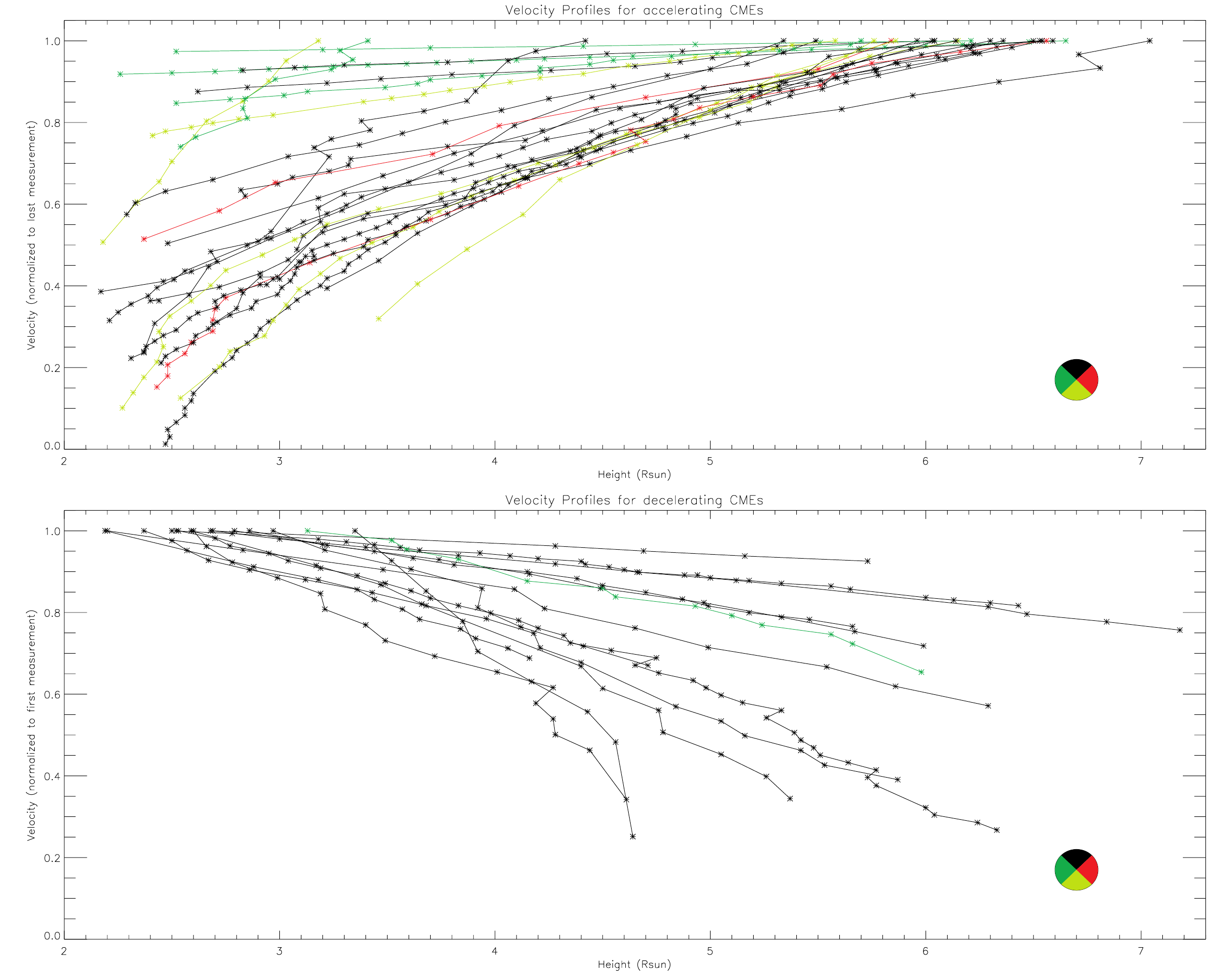}}
\caption{Velocity profiles for accelerating (top) and decelerating (bottom) CMEs without low coronal signatures, as a function of height and normalized to their final (respectively initial) velocity. All profiles are color-coded according to the principal angle of the CME as measured by CACTus.}
\label{velocityprofiles} 
\end{figure}

The velocity profiles for all stealth events in our sample are shown in Figure~\ref{velocityprofiles}, as a function of height and for accelerating and decelerating CMEs separately. For ease of display in a single plot and to facilitate comparison between the velocity profiles, the measurements were normalized with respect to the final CME speed for accelerating CMEs, while for the decelerating ones the initial velocity was used as reference speed. The profiles were colored according to their principal angle of propagation (measured by CACTus). The color code is explained by the schematic sun drawn in the bottom right corner. This figure clearly shows that most stealth CMEs are accelerating and reveals two populations in the top panel. A first group of CMEs is launched at nearly their final speed and accelerates very little, while a second group of events accelerates gradually over the LASCO FOV. Additionally, the bottom panel shows that all but one of the decelerating CMEs originated from the north.

\citet{Sheeley1999} distinguished two CME classes: gradual CMEs that seem to originate from rising prominences and their cavities and have leading edges that accelerate gradually to a velocity in the range of 400 to 600~$\textrm{km s}^{-1}$ within $30 R_\odot$; and impulsive CMEs, often associated with flares and having typical speeds larger than 750~$\textrm{km s}^{-1}$, decelerating as they propagate outwards. Stealth CMEs fit best in the former category of gradual CMEs as they are rather slow events and in most cases their velocity profiles show a moderate acceleration in the LASCO FOV. Obviously, in the case of CMEs without low coronal signatures, there is no associated prominence observed. Likewise \citet{Macqueen1983} reported that flare-associated events generally exhibit higher speeds and little acceleration with height, while in the case of events associated with eruptive filaments lower initial velocities and large accelerations are observed. Unfortunately, in the case of stealth CMEs we are not able to distinguish between these two categories as we do not observe any related flares or filaments.

\subsection{Angular width}
\label{widthsection}

\begin{figure}[H] 
\centerline{\includegraphics[width=1.0\textwidth]{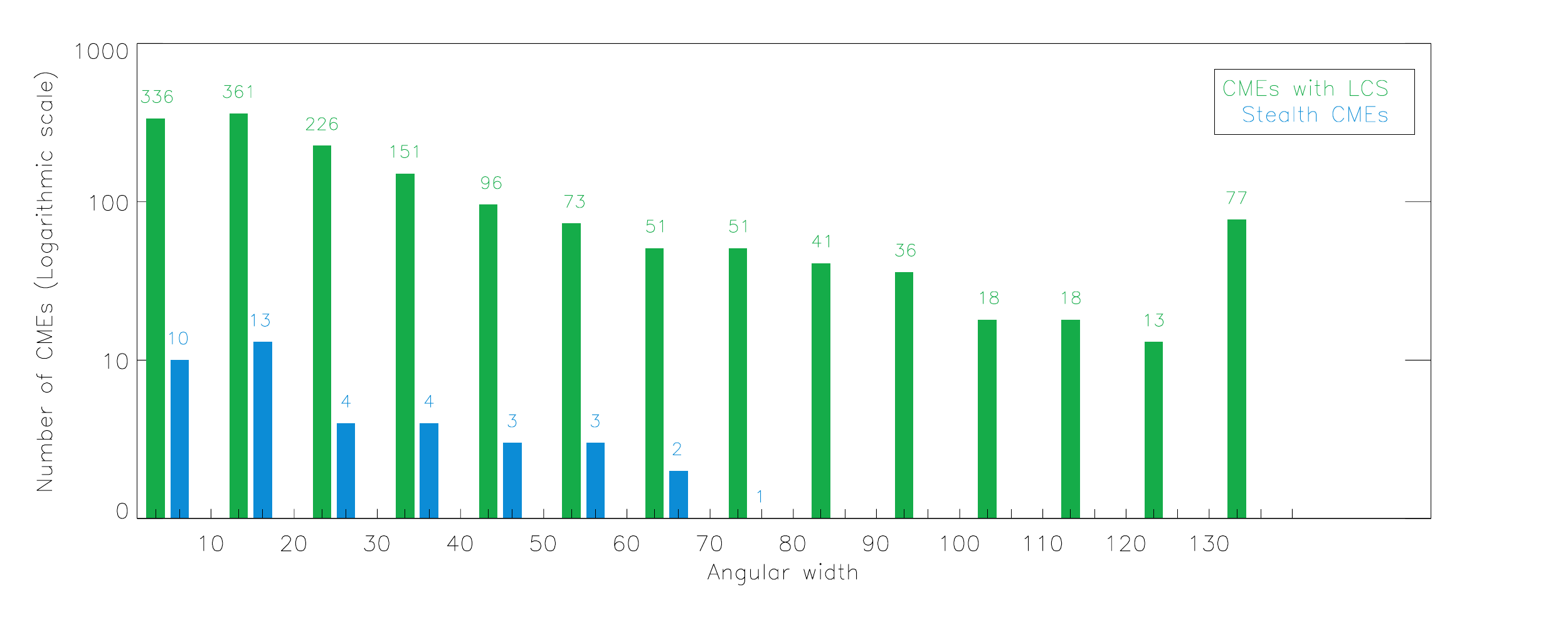}}
\caption{Distribution of the angular width for coronal mass ejections with and without low coronal signatures as measured by the CACTus algorithm and plotted on a logarithmic scale (bin size = 10\degree).}
\label{angularwidth} 
\end{figure}

The angular width distributions for stealth CMEs and CMEs with low coronal signatures are shown in Figure~\ref{angularwidth} on a logarithmic scale. The stealth CMEs show a clear peak around $20\degree$ and the distribution for CMEs with LCS indicates narrow events are more common than large-scale CMEs. A maximum around $20\degree$-$25\degree$ was reported by \citet{Robbrecht2009_cactus} as well when studying the complete CACTus LASCO CME database for solar cycle 23. They compared their results with the manual CDAW CME catalog and noted the latter shows a flatter distribution. This can be explained by the fact that the angular width of a CME is not well defined and large discrepancies are sometimes found when comparing manual and automated measurements, especially for wide CMEs. Additionally, it is known \citep{Robbrecht2004, Yashiro2008} that CACTus detects more narrow CMEs because these narrow events are sometimes regarded as outflows by operators and therefore not recorded as a CME in the CDAW catalog.

To avoid a bias due to these differences in angular width measurements in different catalogs, only the angular widths of CMEs with and without low coronal signatures as measured by CACTus are compared here. While visually both distributions do not seem to differ strongly, Figure~\ref{angularwidth} does suggests that stealth CMEs are comparatively narrow events. The angular width of most stealth CMEs in our sample is below $50\degree$, although CACTus detected some outliers with a much larger width as well. All CMEs with a width larger than 80 degrees were associated with low coronal signatures of an eruption. \citet{Ma2010} report the angular width of their set of stealth CMEs is below $40\degree$. As was the case for the CME velocities, this difference may be explained by their smaller sample size or by the effect of the solar cycle on CME angular width. \citet{Yashiro2004} observed an increase in the average angular CME width from $47\degree$ at the time of solar minimum (1996) to $61\degree$ in the early phase of solar maximum (1999), followed by a decrease to $53\degree$ in 2002, the late phase of solar maximum.

\begin{figure}[H] 
\centerline{\includegraphics[width=1.0\textwidth]{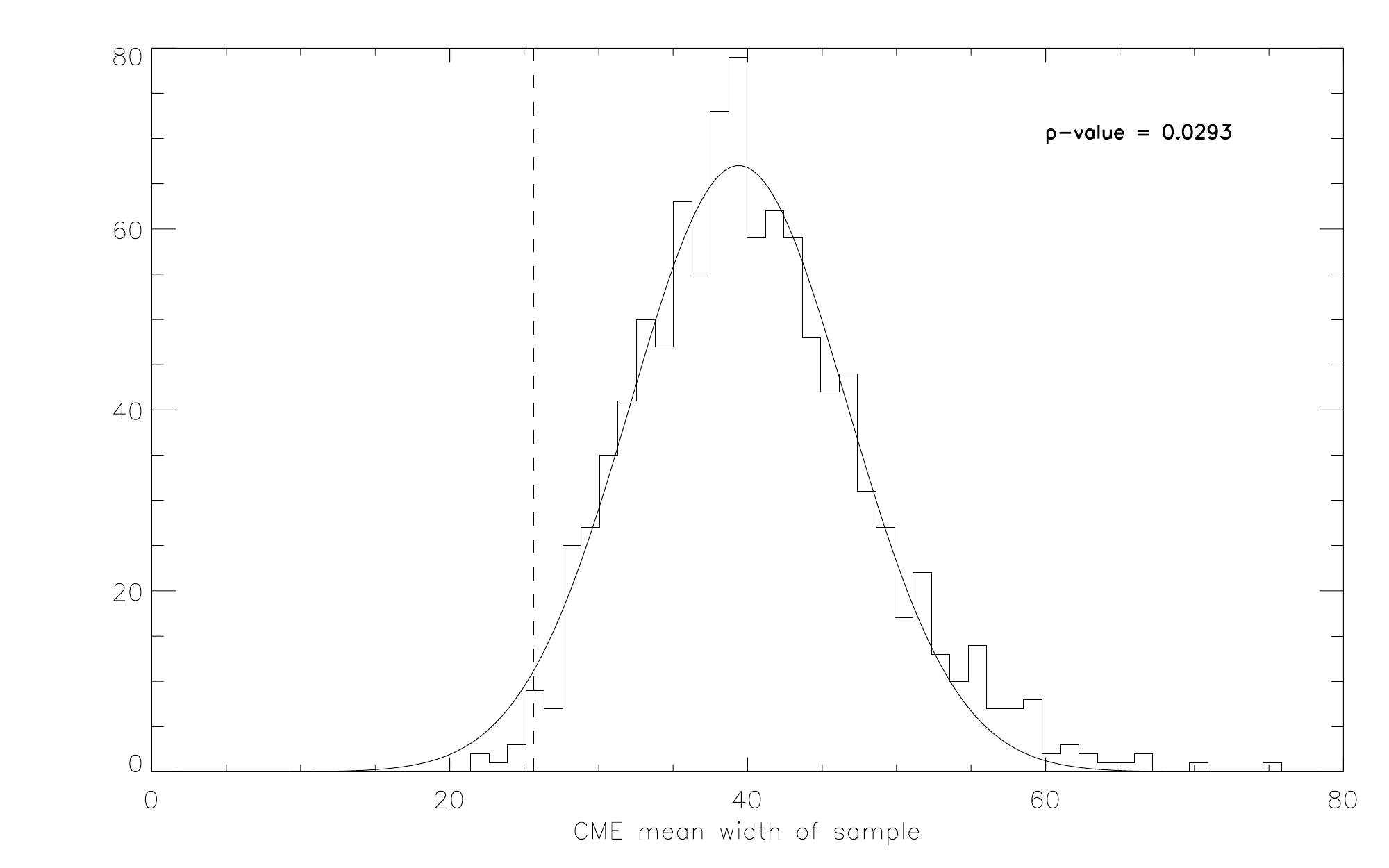}}
\caption{Distribution of the mean angular width of 1000 random samples of 40 coronal mass ejections. The dashed line indicates the mean angular width that was found for the CMEs without low coronal signatures.}
\label{samplewidths} 
\end{figure}

As before, we performed a statistical analysis to evaluate the influence of our small sample size on the angular width distribution for CMEs with low coronal signatures. We computed the mean angular width for 1000 samples of 40 randomly selected CMEs, which resulted in a plot very similar to Figure~\ref{samplevelocity}. In this case, however, the mean widths do not form a true normal distribution. Nevertheless, we have fitted a gaussian distribution with mean $\mu = 39.40$ and standard deviation $\sigma = 7.27$ to this data. As Figure~\ref{samplewidths} illustrates, the actual angular width distribution is well reproduced by the central bell-shape of this gaussian distribution, however the tails do not fit properly. In fact, the left tail, which is of most interest to us since that is where the stealth CME mean angular width value of $25.65$ is found, is overestimated by this fit. The true distribution is lower in the left tail (and higher in the right one). Because the p-value corresponds to the area below the distribution function, the p-value ($p= 0.029$) we find assuming a normal distribution is higher than the true value. Since this p-value is already very low, it is quite likely that the stealth CMEs do not have the same angular width properties as a random sample of 40 events.

\subsection{Scale invariance}

Many authors have searched for a power-law behavior in the frequency distributions of different solar parameters. For example, \citet{Yashiro2006} examined the difference in power-law index for flares with and without an associated coronal mass ejection as a function of different flare parameters (peak flux, fluence, duration). These frequency distributions are often interpreted based on the concept of self-organized criticality \citep[SOC, e.g.][] {Aschwanden2011-1, Aschwanden2011-2}. SOC \citep{Bak1988} describes how dissipative dynamical systems naturally evolve into a minimally stable state through driving by weak external perturbations. A subsequent minor event can then start a chain reaction by which any number of elements in the system may be affected. 

\citet{Lu1991} studied how SOC applies in the solar corona and interpreted a solar flare as an avalanche of many small reconnection events.  The power-law distribution for the occurrence of solar flares is then a direct consequence of the SOC. It also implies that flares are scale-invariant: flares of all sizes are the result of the same physical process and their strength is determined by the number of elementary reconnection events involved. This reasoning can arguably be extended to all solar parameters for which a power-law can be derived.

\begin{figure}[H] 
\centerline{\includegraphics[width=1.0\textwidth]{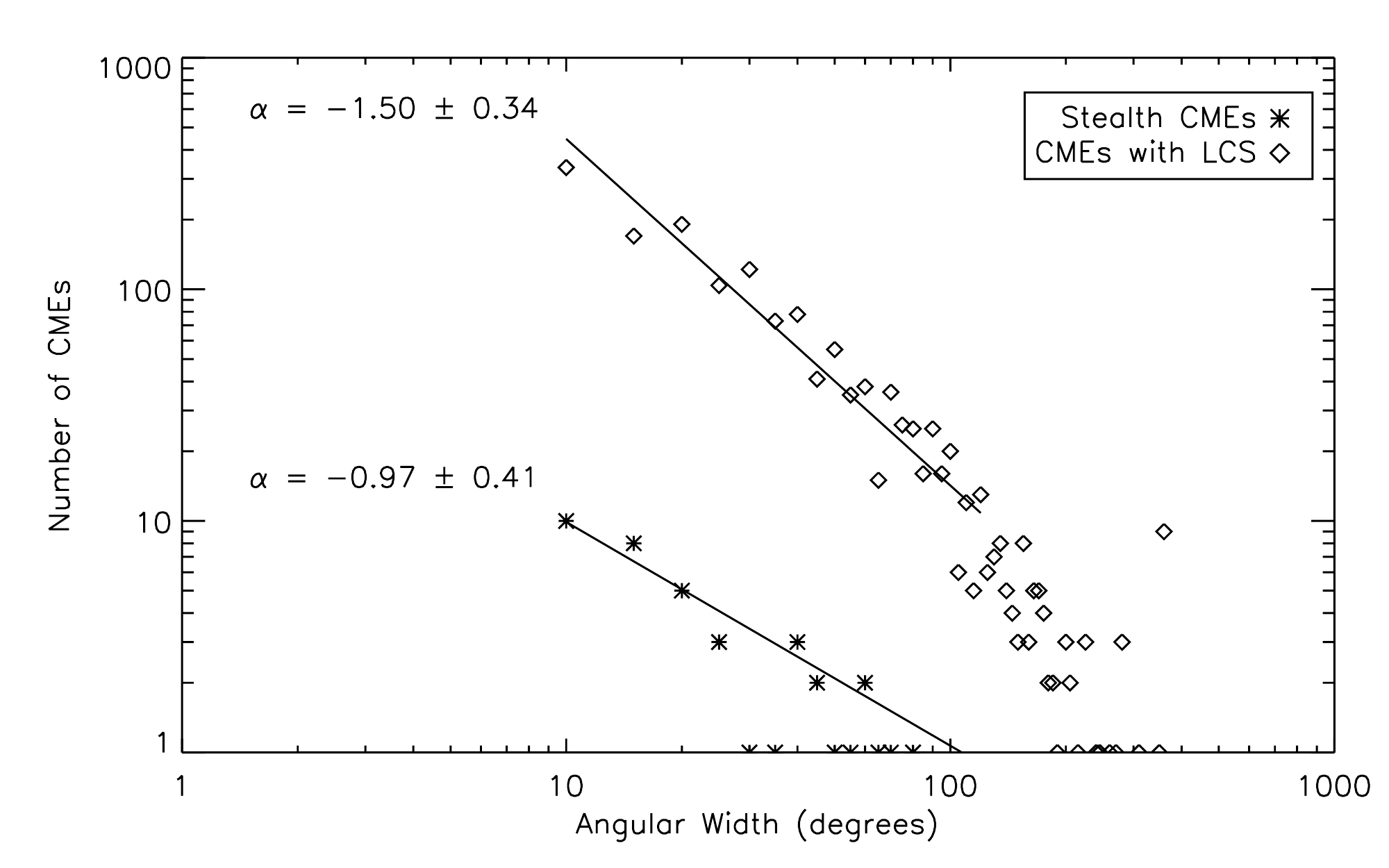}}
\caption{Frequency distributions for the width of coronal mass ejections with (diamonds) and without (asterisk) low coronal signatures as measured by the CACTus algorithm. The linear behaviour on a logarithmic scale implies a power-law that indicates the scale-invariance of CMEs. The stealth CME function exhibits a significantly flatter slope in comparison to CMEs with LCS. The $1\sigma$ uncertainty estimate for $\alpha$  is based on the standard deviation of the angular width distribution.}
\label{scaleinvariance} 
\end{figure}

\citet{Robbrecht2009_cactus} studied CME width histograms for CMEs detected by their CACTus algorithm on a logarithmic scale and found a linear behavior over a large range of angular widths with a slope $\alpha \approx -1.66$. This obtained scale invariance implies that there is no characteristic size for a CME. Figure~\ref{scaleinvariance} shows the frequency distributions for CMEs with and without LCS as a function of width. The distribution for CMEs with low coronal signatures is best described by a linear fit with a slope $\alpha \approx -1.49$, while $\alpha \approx -0.97$ was found for CMEs without. To make these fits, only the events with a width between $5\degree$ and $120\degree$ were used. Wider CMEs were excluded because, due to projection effects, their width measured by CACTus may not correspond well to their true angular width.

We thus find that the angular width distributions for stealth CMEs and CMEs with LCS have a different slope, suggesting a different initiation mechanism may be at work for each class of events. Note that the CACTus width measurements can be an underestimation of the actual angular width, for example in case only part of the CME is detected. For nine out of 40 stealth events, the angular width was underestimated by at least $10\degree$, implying that these CMEs should be counted in a higher histogram bin and that the actual slope for the CMEs without LCS could be somewhat flatter. 

The difference in slope between both CME classes is significant despite the small sample size in the case of stealth CMEs. It is important to note that there is indeed a clear influence of the small sample size on the slope: when 40 CMEs are randomly selected from the CACTus CME list, the resulting slope value is much smaller than the one found for all normal events together. This flatter distribution is not surprising as only a small random selection is made from all CMEs and including a wide CME in such a small sample influences the slope significantly. In reality the angular width distribution of CMEs is dominated by narrow events, which becomes clear when all detections are taken into account. However, the CMEs without low coronal signatures still stand out when compared to random small samples. Selecting 1000 random sets of 40 events yielded a normal distribution of slope values with mean $\mu = -0.55$ and standard deviation $\sigma = 0.21$. For the slope value of the stealth CMEs ($\alpha = -0.97$), we find a probability less than $2\%$ ($p = 0.015$). This implies that the linear fit is much steeper for stealth CMEs than expected  for randomly selected events, indicating that there are more narrow CMEs without low coronal signatures, compared to a random sample of 40 CMEs. This is in agreement with Section~\ref{widthsection}, where we concluded that CMEs without low coronal signatures are generally narrow events.


\section{Discussion: initiation mechanisms for stealth CMEs}

This study was motivated by the question whether CMEs without low coronal signatures are governed by different physical processes than those that do show clear signs of an eruption: do both classes of CMEs have different initiation mechanisms, or are the stealth CMEs simply at the lower energy end of a CME spectrum? In fact, based on their low velocities, gradual acceleration, limited angular width and most importantly the absence of low coronal signatures of eruption, it is likely that stealth CMEs are not very energetic events. Presumably, all available energy goes into expelling the CME, and little is left to leave observable eruption signatures on the solar disk. Additionally, because we do not observe any signatures of magnetic reconnection close to the solar surface, it is highly likely that the crucial reconfiguration of the coronal magnetic field is occurring at higher altitudes where the low density makes the observation of plasma heating challenging. 

If in fact the initiation of CMEs without low coronal signatures occurs at larger heights, this might explain why we observe stealth CMEs to be predominantly narrow and slow. Depending on the surrounding magnetic field, CMEs usually expand and accelerate as they propagate through the corona. In the case of stealth CMEs, assuming the reconnection indeed occurs higher up in the corona, the CME would have less time to expand and speed up before entering the LASCO field-of-view and thus a more narrow and slow event would be observed by CACTus. 

\citet{Pevtsov2012} suggested an explanation for the occurrence of stealth CMEs. These authors studied two erupting filament channels without filament material inside and report these eruptions produced only minor or very gradual changes in the chromosphere and corona. This points to a gradual loss of equilibrium prior to the eruption. In case of the second eruption, the authors suggest the equilibrium was eroded through flux emergence. For instance, \citet{Wang1999} investigated how emerging flux can destabilize a quiescent filament by removing or opening up the magnetic field overlying the filament. They emphasize that, because the magnetic flux emergence can result in a global rearrangement of the magnetic field topology, the source region may be as much as $25\degree$ away from the erupting filament, and therefore may wrongfully appear to be unrelated to the filament eruption. This chain of events where the destabilization of an empty filament channel leads to a CME without LCS was also proposed by \citet{Robbrecht2009_stealth} in their first case-study of a stealth event.

A different stealth CME initiation scenario takes into account that for 29 out of 40 stealth events, another CME was observed preceding the event without low coronal signatures. These preceding eruptions may have destabilized the coronal magnetic field at higher altitude and triggered a stealth event in this way. In case of two eruptions from the same source region, the first eruption may have facilitated the initiation of the stealth event by opening up the overlying magnetic field lines and thus clearing the path for the second eruption. A similar reasoning might explain the fact that many CMEs without low coronal signatures are observed close to coronal holes, where the open field lines might facilitate an eruption.

In the case of multiple eruptions, the first CME may create a pressure imbalance that would cause the trailing stealth CME to be "pulled" out, instead of being launched and driven from below. This chain of events would indeed not leave clear observable traces on the solar surface. 

Another conceivable scenario is described in \citet{Bemporad2012}, where the authors studied two consecutive CMEs, observed on 21 and 22 September 2009 and approximately 7 hours apart. The first CME was caused by a small prominence eruption towards the north, while the second CME could not be associated with a flare, dimming or erupting prominence. Numerical modelling of these events led the authors to conclude that the initiation mechanisms for both CMEs were different. While the first CME was the result of shearing motions, the second one was a mass outflow caused by the rearrangement of the coronal magnetic field after the first eruption. To obtain this second CME, the strength of the global magnetic field is decisive. In a previous simulation by \citet{Zuccarello2012} the same initial magnetic field configuration, boundary conditions and driving mechanism were used. Only the magnetic field strength of the global dipole differed: it had a value of 1.66~G at the poles, compared to 2.2~G for the simulation by \citet{Bemporad2012}. As a result, no second eruption was observed by \citet{Zuccarello2012}.

All observational evidence presented here points to---at most---weak reconnection occurring close to the solar surface during the initiation of stealth CMEs. The models that best fitted their height-time profiles were the magnetic breakout model and MHD instabilities. In these models the eruption is indeed initiated by reconnection higher up in the corona. As the erupting flux rope starts to rise, a current sheet forms below. The reconnection taking place in this current sheet drives the eruption and is the source of a solar flare and other LCS that may be observed. However, in the case of stealth events, this reconnection below is most likely not very powerful, as the CME is not strongly accelerated and no LCS are detected. Some stealth events even showed a very flat velocity profile and thus experienced limited acceleration when propagating in the LASCO FOV, indicating that the driving of the CME had already stopped below $2 R_\odot$. An interesting question to explore might be whether populations of flaring and non-flaring CMEs that do show LCS of eruption bare any similarities to these CMEs without LCS, especially concerning their velocity profiles. However, that analysis is beyond the scope of this paper.

In this study, 40 CMEs without low coronal signatures (LCS), occurring in 2012, were identified. While this is a low number compared to the 1596 CMEs that CACTus detected that year, it is the largest sample of stealth CMEs studied so far. Moreover, this set of stealth CMEs is clearly distinghuishable from a random set of events. The kinematic properties of the CMEs without low coronal signatures were analyzed and compared to those of regular CMEs. We find that stealth CMEs are very diverse in appearance, and tend to originate in the vicinity of the solar north pole. They are generally slow events, showing a gradual propagation in the LASCO field of view, and have a limited angular width. The scale invariance for stealth CMEs reveals a different power-law than for CMEs with clear low coronal signatures of eruption, suggesting an alternative eruption mechanism may be at work for both classes of events. The height-time profiles of stealth eruptions fit both the breakout model and models of ideal MHD instabilities.

Most probably there is not one single initiation mechanism for stealth CMEs. However, the scenarios presented above do show some similarities. Most importantly, the prime reconnection that facilitates or triggers the stealth CME presumably occurs at higher altitude. The erupting flux rope is not expelled forcefully, but is gradually accelerated or even dragged out by the solar wind. As future work, we plan to confirm these assumptions through numerical simulations of a number of CMEs without low coronal signatures that were identified during this study.


\acknowledgements
The authors are grateful to F.P.~Zuccarello for inspiring discussions and thank K.~Bonte for providing the SoFAST catalog data. We thank the anonymous referee for insightful comments that helped us improve this paper. This research was co-funded by a Supplementary Researchers Grant offered by the Belgian Science Policy Office (BELSPO) in the framework of the Scientific Exploitation of PROBA2, the Inter-university Attraction Poles Programme initiated by BELSPO (IAP P7/08 CHARM), and the European UnionÕs Seventh Framework Programme for Research, Technological Development and Demonstration under Grant Agreement No 284461 (Project eHeroes, www.eheroes.eu). E.~D'Huys and D.B.~Seaton also acknowledge support from BELSPO through the ESA-PRODEX program, grant No. 4000103240. This paper uses data from the CACTus CME and SoFAST EUV flare catalogs, generated and maintained by the SIDC at the Royal Observatory of Belgium (www.sidc.be/cactus and www.sidc.be/sofast). PROBA2/SWAP is a project of the Centre Spatial de Liege and the Royal Observatory of Belgium funded by the Belgian Federal Science Policy Office (BELSPO).

{\it Facilities:} \facility{PROBA2}, \facility{SOHO}, \facility{GOES}, \facility{STEREO}, \facility{SDO}


\bibliographystyle{plainnat}
\bibliography{stealth_bibliography} 
\pagebreak
\appendix

\section{Table of CMEs without low coronal signatures}

\begin{deluxetable}{ccccccc}
\tablecolumns{7}
\tablewidth{0pc}
\tablecaption{CACTus detection parameters for the CMEs without low coronal signatures observed in 2012. The first four columns indicate the date and time of each event. The principal angle in the fifth column is expressed in degrees, counting counterclockwards from the north. The angular width and median velocity of the CMEs are given in the last two columns.}
\tablehead{ 
\colhead{Start Date}  & \colhead{Start Time} & \colhead{End Date}  & \colhead{End Time}  &
\colhead{Principal}  & \colhead{Angular} & \colhead{Median}  \\  
  &  &   &   & \colhead{Angle (\degree)}  & \colhead{Width (\degree)} & \colhead{Velocity ($\textrm{km s}^{-1}$)}
}
\startdata
07 Jan 2012 & 15:24:05 & 07 Jan 2012 & 17:36:05 & 30 & 6 & 431 \\
07 Jan 2012 & 23:48:06 & 08 Jan 2012 & 01:12:06 & 3 & 12 & 142 \\
19 Jan 2012 & 22:36:05 & 20 Jan 2012 & 00:00:06 & 357 & 12 & 856 \\
20 Jan 2012 & 00:24:05 & 20 Jan 2012 & 02:48:05 & 334 & 22 & 390 \\
20 Jan 2012 & 17:12:06 & 20 Jan 2012 & 18:48:07 & 341 & 8 & 418 \\ 
26 Jan 2012 & 16:38:06 & 26 Jan 2012 & 18:48:05 & 20 & 46 & 249 \\
28 Jan 2012 & 04:12:05 & 28 Jan 2012 & 05:00:06 & 311 & 12 & 749 \\
04 Feb 2012 & 09:24:06 & 04 Feb 2012 & 14:48:05 & 357 & 80 & 216 \\
22 Feb 2012 & 12:48:05 & 22 Feb 2012 & 12:48:05 & 20 & 10 & 138 \\
22 Feb 2012 & 23:48:06 & 23 Feb 2012 & 02:00:05 & 9 & 44 & 143 \\
23 Feb 2012 & 12:24:05 & 23 Feb 2012 & 14:48:06 & 45 & 12 & 139 \\
29 Feb 2012 & 19:48:07 & 01 Mar 2012 & 01:36:22 & 16 & 58 & 254 \\
21 Mar 2012 & 23:12:10 & 22 Mar 2012 & 04:12:05 & 168 & 42 & 169 \\
19 Apr 2012 & 01:36:05 & 19 Apr 2012 & 01:36:05 & 176 & 14 & 116 \\
16 May 2012 & 03:24:05 & 16 May 2012 & 08:24:05 & 179 & 64 & 207 \\
03 Jun 2012 & 07:12:05 & 03 Jun 2012 & 12:00:07 & 345 & 36 & 261 \\
09 Jun 2012 & 07:24:05 & 09 Jun 2012 & 08:48:05 & 27 & 20 & 330 \\
17 Jun 2012 & 04:48:05 & 17 Jun 2012 & 07:24:05 & 83 & 12 & 466 \\
07 Jul 2012 & 18:00:06 & 07 Jul 2012 & 22:36:06 & 100 & 14 & 323 \\
13 Jul 2012 & 05:05:54 & 13 Jul 2012 & 08:24:06 & 330 & 18 & 329 \\
14 Jul 2012 & 19:24:06 & 14 Jul 2012 & 19:24:06 & 334 & 10 & 310 \\
17 Jul 2012 & 23:24:06 & 18 Jul 2012 & 01:26:17 & 318 & 18 & 1117 \\
21 Jul 2012 & 02:36:06 & 21 Jul 2012 & 05:00:07 & 76 & 18 & 771 \\
28 Jul 2012 & 14:24:07 & 28 Jul 2012 & 16:12:06 & 338 & 6 & 330 \\
12 Aug 2012 & 20:24:07 & 12 Aug 2012 & 20:48:06 & 338 & 38 & 197 \\
16 Aug 2012 & 05:00:06 & 16 Aug 2012 & 05:48:06 & 40 & 70 & 137 \\
04 Sep 2012 & 03:48:06 & 04 Sep 2012 & 05:24:06 & 344 & 18 & 443 \\
18 Sep 2012 & 02:12:09 & 18 Sep 2012 & 05:12:08 & 3 & 56 & 469 \\
22 Sep 2012 & 07:00:06 & 22 Sep 2012 & 07:12:07 & 178 & 22 & 136 \\
20 Oct 2012 & 23:48:06 & 21 Oct 2012 & 00:36:08 & 15 & 32 & 138 \\
28 Oct 2012 & 01:48:07 & 28 oct 2012 & 02:24:07 & 115 & 8 & 336 \\
03 Nov 2012 & 23:24:07 & 04 Nov 2012 & 01:48:06 & 346 & 6 & 257 \\
14 Nov 2012 & 00:00:06 & 14 Nov 2012 & 07:00:06 & 242 & 54 & 231 \\
16 Nov 2012 & 13:36:31 & 16 Nov 2012 & 16:00:06 & 353 & 8 & 292 \\ 
25 Nov 2012 & 18:36:06 & 25 Nov 2012 & 18:36:06 & 24 & 22 & 136 \\
17 Dec 2012 & 03:36:07 & 17 Dec 2012 & 03:36:07 & 156 & 10 & 312 \\
18 Dec 2012 & 08:24:06 & 18 Dec 2012 & 09:12:10 & 334 & 38 & 262 \\
18 Dec 2012 & 18:36:06 & 18 Dec 2012 & 19:12:07 & 340 & 14 & 138 \\
19 Dec 2012 & 18:36:06 & 19 Dec 2012 & 20:48:06 & 350 & 10 & 364 \\
20 Dec 2012 & 21:17:39 & 20 Dec 2012 & 23:48:06 & 12 & 26 & 277
\enddata
\label{stealthtable}
\end{deluxetable}

\end{document}